\documentclass[aps,showpacs,amsfonts,nofootinbib,preprintnumbers,nobalancelastpage]{revtex4}
\usepackage{graphicx}
\usepackage{epsfig}
\usepackage{xcolor}
\usepackage{rotating}
\usepackage{amsmath}
\usepackage{braket}
\usepackage{multirow}      
\usepackage{booktabs}      
\usepackage{array}         

\newcommand{\Dfb}{\mbox{$\raisebox{2mm}{\boldmath ${}^\leftrightarrow$}
\hspace{-4mm} D$}}
\newcommand{\Dfba}{\mbox{$\raisebox{2mm}{\boldmath ${}^\leftrightarrow$}\hspace{-4mm} D^a$}}

\begin{document}


\title{Impact of fermionic operators on the Higgs width measurement}

\author{Eduardo da Silva Almeida}
\email{eduardo.silva.almeida@usp.br}
\affiliation{Instituto de F\'{\i}sica,
             Universidade de S\~ao Paulo, S\~ao Paulo -- SP,
             05508-090, Brazil.}

\author{O.\ J.\ P.\ \'Eboli}
\email{eboli@if.usp.br}
\affiliation{Instituto de F\'{\i}sica,
             Universidade de S\~ao Paulo, S\~ao Paulo -- SP,
             05508-090, Brazil.}

\author{M.\ C.\ Gonzalez--Garcia} \email{concha@insti.physics.sunysb.edu}
\affiliation{%
  Instituci\'o Catalana de Recerca i Estudis Avan\c{c}ats (ICREA),}
\affiliation {Departament d'Estructura i Constituents de la Mat\`eria, 
Universitat
  de Barcelona, 647 Diagonal, E-08028 Barcelona, Spain}
\affiliation{%
  C.N.~Yang Institute for Theoretical Physics, SUNY at Stony Brook,
  Stony Brook, NY 11794-3840, USA}

\begin{abstract}

  The off-shell Higgs production in $p p \to ZZ$ at LHC provides at
  present the most direct measurement of the Higgs width in the
  absence of beyond the standard model contributions. Here, we analyze
  the impact of anomalous $Z$ couplings to fermions on the Higgs width
  determination.  We show that, despite these couplings being strongly
  constrained by the available electroweak precision data, they can
  substantially affect the Higgs width determination at the LHC Runs 2
  and 3. Conversely, in larger integrated luminosities runs, such as
  those foreseen at the high luminosity LHC and high energy LHC
  setups, the effect of such anomalous interactions in the Higgs width
  measurement is minimal.
  
\end{abstract}

\preprint{YITP-SB-2020-23}

 \maketitle
\renewcommand{\baselinestretch}{1.15}
%

\section{Introduction}

In the Standard Model (SM), the Higgs boson is a relatively narrow
resonance with a width $\Gamma_H^{SM} \simeq 4.1$ MeV which is difficult to
measure directly at the Large Hadron Collider (LHC). Notwithstanding,
as first pointed out in Ref.~\cite{Caola:2013yja}, it is possible to
extract the Higgs width from the study of the off-shell Higgs
production in $ p p \to Z Z$ assuming that this process receives no
contribution from new physics.  This procedure is, in fact, the one
being used by the LHC experimental collaborations to perform such
measurement~\cite{ Aad:2015xua, Aaboud:2018puo, Khachatryan:2016ctc,
  Sirunyan:2019twz}. \smallskip

One characteristic feature of the off-shell Higgs production
$ p p \to Z Z$, is that the Higgs contribution ($g g \to H \to ZZ$)
interferes destructively with the continuum $g g \to Z Z$ generated
via box diagrams ~\cite{Passarino:2012ri,Kauer:2012hd,
  Campbell:2013una}.  This cancellation is affected if new physics
contributes to the off-shell Higgs production, either by altering the
Higgs couplings or by introducing new particles running in the loops
or being exchanged. This fact has been exploited in the literature to
study departures of the Higgs couplings from their SM value, as well
for signals of new physics~\cite{Gainer:2014hha, Englert:2014aca,
  Cacciapaglia:2014rla,Azatov:2014jga, Englert:2014ffa,
  Buschmann:2014sia, Brivio:2014pfa,Corbett:2015ksa, Azatov:2016xik,
  Goncalves:2018pkt,He:2019kgh}. In fact, the off-shell Higgs
production data is already being used by the experimental
collaborations to probe the $HVV$ coupling and to access the impact of
such anomalous Higgs interactions in the determination of its
width~\cite{Sirunyan:2019twz}.  \smallskip

It is also a fact that the off-shell Higgs production possesses a
large background stemming from $ q \bar{q} \to Z Z$ that is much
larger than the off-shell Higgs signal. Consequently, changes in this
background process induced, for example, by anomalous $Z\bar{q}q$
interactions have the potential to affect the determination of the
Higgs width and of possible anomalous Higgs interactions.  This
reaction is similar to $q \bar{q} \to W^+ W^-/W^\pm Z$ used to probe
the triple electroweak gauge couplings (TGC). And as it has been
shown~\cite{ Zhang:2016zsp,Alves:2018nof,
  Almeida:2018cld,Azzi:2019yne} that, despite the precise
determination of the electroweak vector boson couplings to fermions
from electroweak precision data (EWPD), anomalous fermionic couplings
can impact the determination of TGC due to the present and future
precisions of the LHC. It is then a matter of concern, the possible
impact of $Z\bar{q}q$ anomalous interactions in the present and new
future determinations of the Higgs width through the study of the
off-shell Higgs production. It is the goal of this work to quantify
such impact.  \smallskip

To this end, in this work we scrutinize the determination of the Higgs
width through off-shell Higgs production taking into account anomalous
$Z\bar{q}q$ couplings by analyzing the process
\begin{equation}
  p p \to \ell^+ \ell^- \ell^+ \ell^-
\label{eq:proc}
\end{equation}
where $\ell$ stands for $e$ and $\mu$.  We perform our analyses for
the LHC Runs 2 and 3, as well as, for the high-luminosity LHC (HL-LHC)
and high-energy LHC (HE-LHC). Our results show that the presence of
anomalous $Z\bar{q}q$ contributions to Eq.~\eqref{eq:proc} which are
not accounted for in the model to be fitted, translates both in
changes in the apparent precision in the Higgs width determination in
the LHC Runs 2 and 3, as well as into a shift in its derived central
value.  Conversely, the expected larger integrated luminosities at the
HL-LHC and HE-LHC runs should be enough to mitigate these effects,
making the determination of the Higgs width in those setups robust
under the presence of anomalous $Z$ couplings within its present
bounds. \smallskip

\section{Analyses framework}
\label{sec:forma}

We parametrize the effects of new physics on the $Z$ couplings to
quarks at low energies by dimension-six effective operators where the
$SU(2)_L \otimes U(1)_Y$ gauge symmetry is linearly
realized~\cite{Buchmuller:1985jz, Leung:1984ni, DeRujula:1991ufe,
  Hagiwara:1993ck, GonzalezGarcia:1999fq, Grzadkowski:2010es,
  Passarino:2012cb}, that is,
\begin{equation}
{\cal L}_{\rm eff} = {\cal L}_{\text SM} + \sum_i
\frac{f^{(i)}_i}{\Lambda^{2}} {\cal O}^{(6)}_i \;\; ,
\label{l:eff}
\end{equation}
where the dimension-six operators ${\cal O}^{(6)}_i$ involve gauge
bosons, Higgs doublets, fermionic fields, and covariant derivatives of
these fields. Here, we considered the following operators that modify
the left- and right-handed couplings of the $Z$ to quarks
\begin{equation}
 \begin{array}{l@{\hspace{1cm}}l@{\hspace{1cm}}l}
  & 
{\cal O}^{(1)}_{\phi Q,ij}=
\phi^\dagger (i\,\Dfb_\mu \phi)  
(\bar Q_i\gamma^\mu Q_{j}) \;\;,
& 
{\cal O}^{(3)}_{\phi Q,ij}=\phi^\dagger (i\,{\Dfba}_{\!\!\mu} \phi) 
(\bar Q_i\gamma^\mu T_a Q_j) \;\;,
\\
&
&
\\
& {\cal O}^{(1)}_{\phi u,ij}=\phi^\dagger (i\,\Dfb_\mu \phi) 
(\bar u_{R_i}\gamma^\mu u_{R_j}) \;\;,
& {\cal O}^{(1)}_{\phi d,ij}=\phi^\dagger (i\,\Dfb_\mu \phi) 
(\bar d_{R_i}\gamma^\mu d_{R_j}) \;\;,
\end{array}
\label{eq:hffop}
\end{equation}
where we defined $\tilde \phi=i \sigma_2\phi^*$,
$\phi^\dagger\Dfb_\mu\phi= \phi^\dagger D_\mu\phi-(D_\mu\phi)^\dagger
\phi$ and
$\phi^\dagger \Dfba_{\!\!\mu} \phi= \phi^\dagger T^a D_\mu
\phi-(D_\mu\phi)^\dagger T^a \phi$ with $T^a=\sigma^a/2$ and
$\sigma^a$ standing for the Pauli matrices.  We have also used the
notation of $Q$ for the quark doublet and $f_R$ for the $SU(2)_L$
singlet quarks. Here, $i, j$ are family indices and, for the sake of
simplicity, we consider only diagonal family couplings that are
generation independent. \smallskip

In addition to the above anomalous couplings, we also considered the
electroweak dipole operators
\begin{equation}
\begin{array}{l@{\hspace{1cm}}l@{\hspace{1cm}}l}
{\cal O}_{uW,ij} =  i \overline{Q}_i \sigma^{\mu\nu} u_{R,j} \widehat{W}_{\mu\nu}
  \tilde\phi \;\;\;, 
  &{\cal O}_{uB,ij}  = i \overline{Q}_i \sigma^{\mu\nu} u_{R,j} \widehat{B}_{\mu\nu}
  \tilde\phi \;\;\;, 
\\
{\cal O}_{dW,ij} = i \overline{Q}_i \sigma^{\mu\nu} d_{R,j} \widehat{W}_{\mu\nu}
  \phi \;\;\;, 
& {\cal O}_{dB,ij} = i\overline{Q}_i \sigma^{\mu\nu} u_{R,j} \widehat{B}_{\mu\nu}
  \phi \;\;\;,
\end{array}
\label{eq:dipole}
\end{equation}
where we defined
$\widehat{B}_{\mu\nu} \equiv i(g^\prime/2) B_{\mu\nu}$ and
$\widehat{W}_{\mu\nu} \equiv i(g/2) \sigma^a W^a_{\mu\nu}$, with $g$
and $g^\prime$ being the $SU(2)_L$ and $U(1)_Y$ gauge couplings
respectively. \smallskip

\begin{table}
\begin{tabular}{|c|c|c|}
\hline
    operator  & EWPD  & EWPD+EWDBD
\\
\hline
${\cal O}^{(1)}_{\phi Q}$  & [-0.083,0.10] & [-0.034,0.11]
  \\
  \hline
  ${\cal O}^{(3)}_{\phi Q}$
  & [-0.60,0.12] & [-0.45,0.13]
  \\
  \hline
  ${\cal O}^{(1)}_{\phi d}$  & [-1.2,-0.13] & [-0.64,-0.007]
  \\
  \hline
  ${\cal O}^{(1)}_{\phi u}$  & [-0.25,0.37] & [-0.17,0.37]
  \\
  \hline
  ${\cal O}_{uW} $  & [-10.,10.] & [-0.29,0.29]
  \\
  \hline
  ${\cal O}_{uB} $  & [-41.,41.] & [-1.9,1.9]
  \\
  \hline
  ${\cal O}_{dW} $  & [-10.,10.] & [-0.36,0.36]
  \\
  \hline
  ${\cal O}_{dB} $  & [-38.,38.] & [-1.9,1.9]
  \\
  \hline
\end{tabular}
\caption{95\% C.L. allowed regions for the Wilson coefficients
  $f/\Lambda^2$ in TeV$^{-2}$ taking into account the electroweak
  precision data and diboson
  production~\cite{Almeida:2018cld,Almeida:2019jqy}.}
\label{tab:limits}
\end{table}

Presently, the Wilson coefficients of the dimension-six operators in
Eqs.~\eqref{eq:hffop} and \eqref{eq:dipole} are bound by EWPD and
electroweak diboson data (EWDBD)~\cite{Almeida:2018cld,deBlas:2017wmn,
  Falkowski:2019hvp,Almeida:2019jqy, Dawson:2020oco}.  For
convenience, in Table~\ref{tab:limits}, we summarize these limits as
derived in Refs.~\cite{Almeida:2018cld,Almeida:2019jqy}. As seen in
this table they are roughly of the order of 1 TeV$^{-2}$. \smallskip

Generically, in order to extract the Higgs width from off-shell Higgs data,
the gluon-gluon cross section is parametrized as
\begin{equation}
  \sigma( g g \to \ell^+ \ell^- \ell^+ \ell^-) =
  \sigma_{\rm cont} + \sqrt{X} \sigma_{\rm inter} + X \sigma_{H}  \;,
\label{eq:X}
\end{equation}  
where the coefficient
\begin{equation}
  X = \mu_{4\ell} \times \frac{\Gamma_H}{\Gamma_H^{SM}}
\label{eq:Xdef}
\end{equation}
is the variable to be determined from the analysis. 
$\Gamma_H$ and $\Gamma_H^{SM}$ stand for the Higgs total width to be
determined and its SM predicted value respectively.  $\mu_{4\ell}$ is
the signal strength for the Higgs production in the four-lepton
channel which can be independently constrained by on-shell Higgs
data. \smallskip

We denoted by $\sigma_{\rm cont}$ the continuum cross section, by
$\sigma_H$ the Higgs contribution and by $\sigma_{\rm inter}$ the
interference between continuum and Higgs contributions. In our
simulations these gluon-gluon initiated contributions were evaluated
using MCFM-9.1~\cite{Campbell:2013una}.  In our calculations, we used
the NNPDF 3.0 parton distribution functions~\cite{Ball:2011uy} in
LHAPDF 6.2.3~\cite{Buckley:2014ana}. \smallskip

An irreducible background to the above process is induced at tree
level by
\begin{equation}
q\,\bar q \rightarrow \ell^+ \ell^- \ell^+ \ell^-\;.
\label{eq:qqzz}
\end{equation}
In order to evaluate the contributions of the above anomalous
$Z \bar{q}q$ couplings to the process in Eq.~\eqref{eq:qqzz}, we used
the package FeynRules~\cite{Alloul:2013bka} to generate the UFO files
needed as input into MadGraph 5~\cite{Alwall:2011uj}, that we employed
to compute the tree level cross sections.\smallskip

Finally, let us notice that the anomalous $Z\bar{q}q$
couplings also
contribute to the box diagram $g g \to ZZ$, therefore, it modifies the
continuum and interference cross sections.
Indeed the anomalous $Z\bar{t}t$ anomalous coupling contributions have
already been considered when studying new-physics effects in off-shell
Higgs production; see for instance Ref.~\cite{Azatov:2014jga}.
However, the effect os anomalous $Z$ couplings to the light quarks has
not been analyzed in the $q\bar{q}$ process, neither in the
gluon-gluon loops.
We took into account this effect for the couplings in
Eq.~\eqref{eq:hffop} to the lowest order in $1/\Lambda^2$ and it was
evaluated using a modified version of MCFM-9.1.\smallskip

\section{Results}
\label{sec:results}

In what follows we are going to use as main observable the four-lepton
invariant mass distribution of events. To do so, we performed a parton
level simulation and we imposed cuts similar to the ones in
Ref.~\cite{Sirunyan:2019twz}, that is, we required the final state
leptons to satisfy
\begin{equation}
  p_T^\ell > 10 \hbox{ GeV } \;\;\;\;,\;\;\;\;
  |\eta_\ell| < 2.4 \;\;\;\;,\;\;\;\;
  p_{T, {\rm hardest}}^{ \ell} > 20 \hbox{ GeV} 
\end{equation}
and that the events should exhibit two same-flavor opposite charge
lepton pair close to the $Z$ mass, {\em i.e.}
\begin{equation}
  40 < m_{\ell\ell} < 120 \hbox{ GeV} \; .
\end{equation}
We analyzed four setups: for the LHC Run2 (Run3/HL) we considered a
center-of-mass energy of 13 (14) TeV and an integrated luminosity of
140 (400/3000) fb$^{-1}$, while for the HE-LHC we assumed a
center-of-mass energy of 27 TeV and an integrated luminosity of 15
ab$^{-1}$. \smallskip

\begin{figure}[h!]
\centering
 \includegraphics[width=0.7\textwidth]{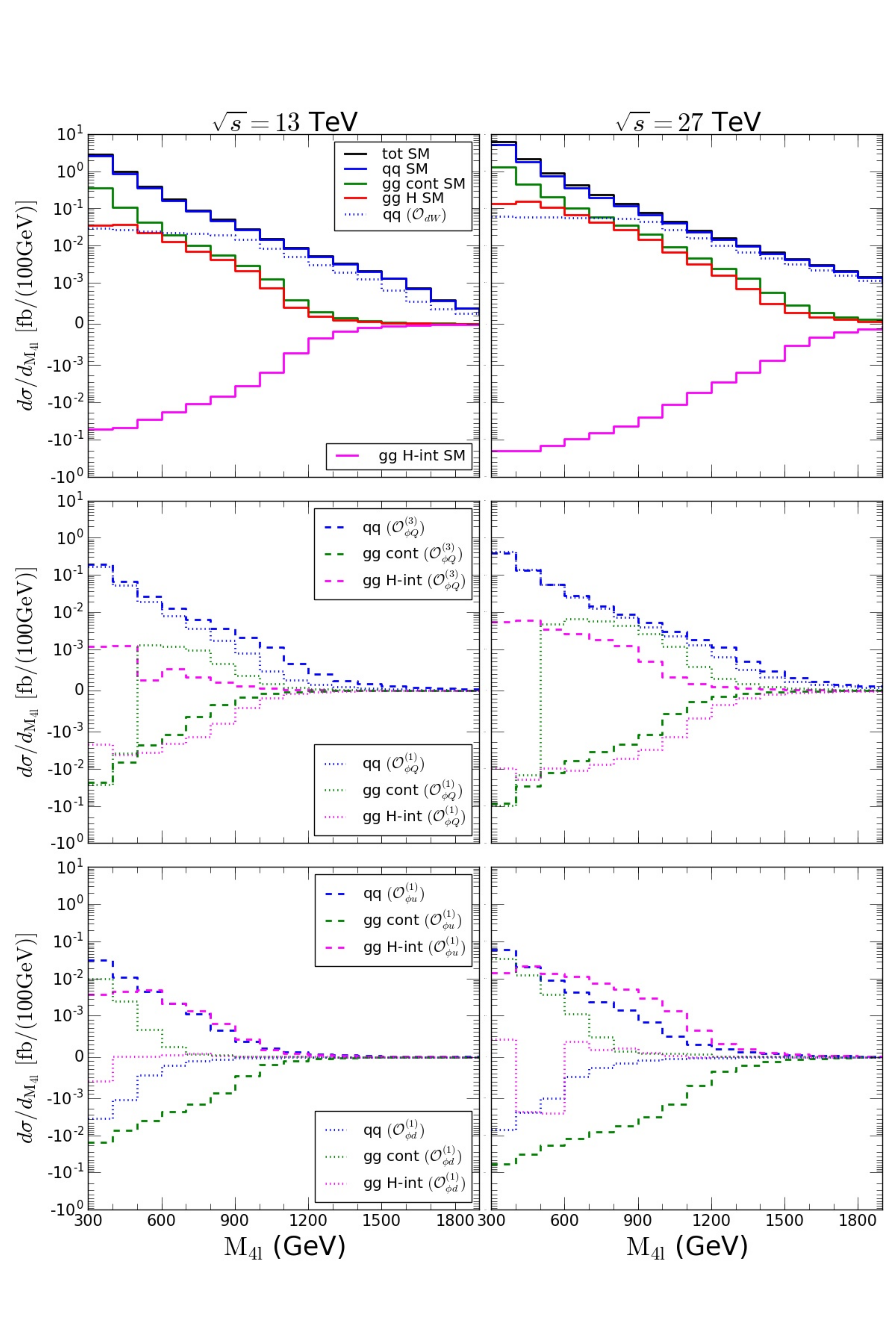}
 \caption{Four-lepton invariant mass distributions for the different
   processes relevant to our study as labeled in the figure. All
   anomalous contributions have been computed for a Wilson coefficient
   $f/\Lambda^2=1$ TeV$^{-2}$.}
  \label{fig:m4l}
\end{figure}

In Figure~\ref{fig:m4l}, we depict the resulting four-lepton
invariant mass distributions for the center-of-mass energies of 13 TeV
and 27 TeV; the results for 14 TeV are similar to the 13 TeV ones.
In the top two panels we display the SM expected distributions, as
well as the quark-antiquark fusion contribution from the dipole
operator ${\cal O}_{dW}$ (more below). In the middle and lower panels
of Fig.~\ref{fig:m4l}, we present the $1/\Lambda^2$ anomalous
contributions of the operators in Eq.~\eqref{eq:hffop} for the quark-
and gluon-fusion channels.  All the anomalous cross sections shown
have been evaluated for a Wilson coefficient $f/\Lambda^2=1$
TeV$^{-2}$.  Amplitudes induced by the operators in
Eq.~\eqref{eq:hffop} interfere with the SM ones and the cross sections
have been evaluated to linear order in the corresponding Wilson
coefficient.  Indeed, within the present bounds in Table
~\ref{tab:limits} this is the dominant contribution. On the contrary,
amplitudes induced by dipole operators do not interfere with the SM
one and therefore the cross sections shown are quadratic in the
corresponding Wilson coefficient~\footnote{Notice that we have only
  considered the effects of the dipole operators to quark-fusion
  processes since their effects in gluon initiated processes are not
  available yet.}. \smallskip

From this figure we see that quark-antiquark fusion cross sections
induced by some of the operators in Eq.~\eqref{eq:hffop} are
quantitatively comparable in size and shape to those of the SM Higgs
induced subprocesses.  Consequently, if these contributions are
present in the data but not included in the analysis they can alter
the extracted value of the $X$ coefficient in Eq.~\eqref{eq:Xdef}.
Quantitatively, within the present bounds on the Wilson coefficients,
we find that the largest potential effect would correspond to
${\cal O}^{(3)}_{\phi Q}$ in which we focus in the
following.\smallskip

Conversely, from the upper panels of Fig.~\ref{fig:m4l} we learn that
anomalous dipole contributions exhibit a spectrum different than the SM
one with the dipole contribution becoming more important at high
four-lepton invariant masses. This is expected since dipole scattering
amplitude grows as $m_{4\ell}^2$ at high
energies~\cite{Corbett:2017qgl}. This, in principle, makes the effect
of these operators easier to disentangle from a shift in the Higgs
width. However, as we will show below, this relies on having enough
statistics in the larger invariant mass bins. \smallskip

\begin{figure}[h!]
\centering
 \includegraphics[width=0.9\textwidth]{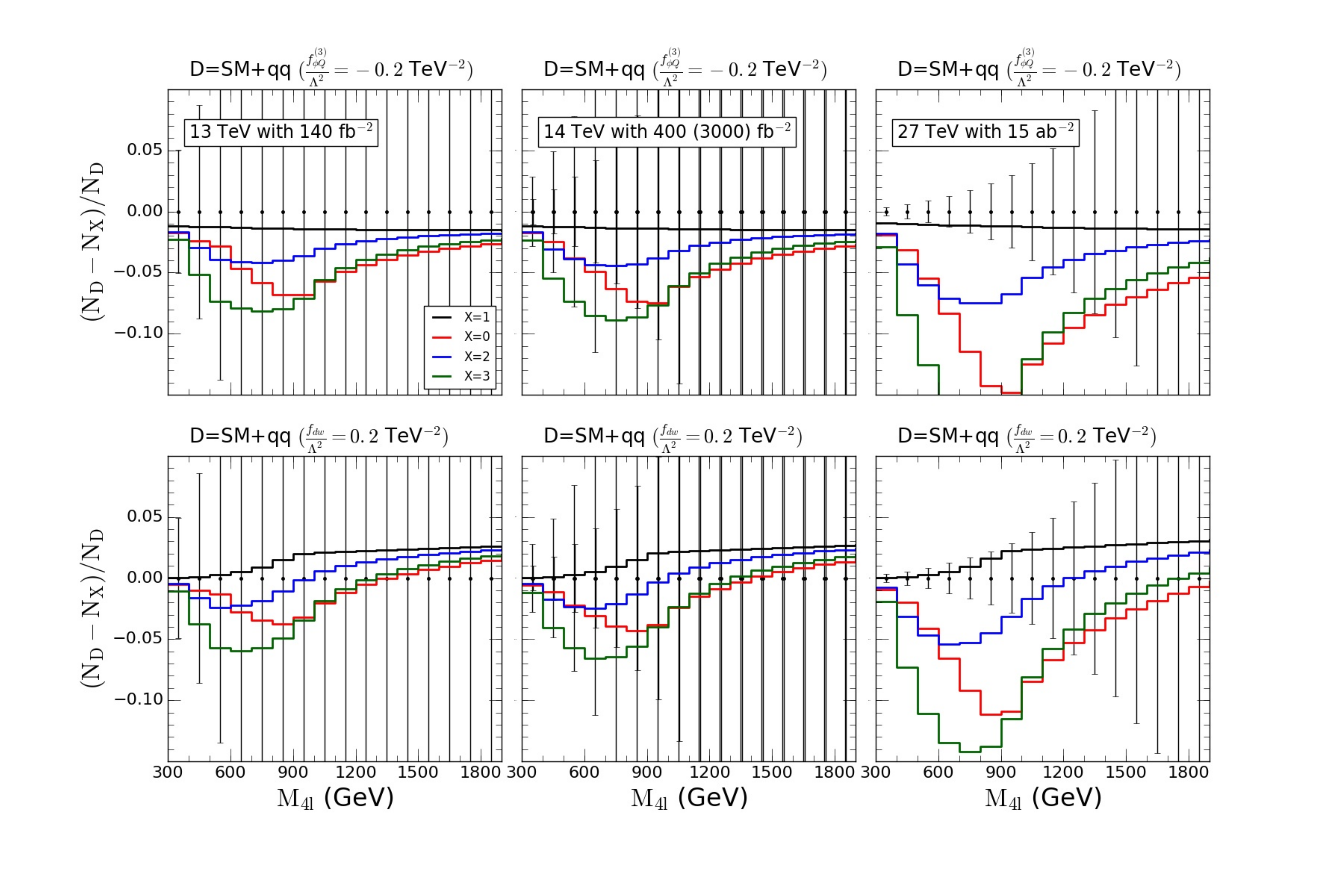}
 \caption{Relative variation of the expected number of event
   distribution as induced by anomalous $Z\bar{q}q$ couplings versus a
   non-standard Higgs width coefficient $X$. In the upper (lower)
   panels $f_{\phi Q}^{(3)}/\Lambda^2=-0.2$ TeV$^2$
   ($f_{dW}/\Lambda^2=0.2$ TeV$^2$).  The left, middle, and right
   panels corresponds to the LHC Run2, Run3 /HL, and HE-LHC
   respectively.  For comparison, we show as error bars the
   corresponding statistical uncertainties for an hypothetical
   observation according to $N_D$. In the central panel the larger
   (smaller) error bars correspond to ${\cal L}= 400\;(3000)$
   fb$^{-1}$.}
\label{fig:errors}
\end{figure}

To further illustrate and quantify the impact of the anomalous
$Z\bar{q}q$ couplings in the Higgs width determination, we show in
Figure~\ref{fig:errors} the relative change in the number of events
induced by the anomalous coupling compared to that induced by a shift
of the $X$ coefficient
\[
  \frac{N_D - N_X}{N_D} \;,
\]
where by $N_D$ we denote the expected number of events considering the
SM $q\bar{q}$ and $gg$ continuum contributions, as well as the one due
to the $Z$ anomalous coupling of coefficient $f$, this is
\begin{equation}
  N_D = {\cal L} \times\big[\sigma^{SM}_{q\bar{q}} + \sigma_{q
      \bar{q}}^{ano}(f) 
    + \sigma_{g g} (X=1, f) \big] \;,
\label{eq:ndata}
\end{equation}
where ${\cal L}$ is the integrated luminosity.  $N_X$ stands for the
number of events expected from SM $q\bar{q}$ and $gg$ continuum
contributions together with the Higgs one with a different value of
the coefficient $X$; see Eq.~\eqref{eq:X}.
\begin{equation}
  N_X = {\cal L} \times\big[\sigma^{SM}_{q\bar{q}}
    + \sigma_{g g} (X,f=0) \big] \;.
\label{eq:nx}
\end{equation}
The upper panels correspond to $f=f_{\phi Q}^{(3)}/\Lambda^2=-0.2$
TeV$^2$ and the lower panels to $f_{dW}/\Lambda^2=0.2$ TeV$^2$, values
well within their presently allowed 2$\sigma$ range.  The left, middle
and right panels correspond to the LHC Run2, Run3 /HL, and HE-LHC
respectively.  For comparison we show as error bars the corresponding
statistical uncertainties for an hypothetical observation according to
$N_D$. \smallskip

As we can see, the largest statistical weight originates from
invariant-mass bins smaller than 500 (700/900/1200) GeV for the LHC
Run 2 (LHC Run 3/HL-LHC/HE-LHC).  It is interesting to notice that the
growth of the anomalous cross sections at high four-lepton invariant
masses for the dipole operator ${\cal O}_{dW}^{(3)}$ turns out
not to be very statistically significant. \smallskip

Altogether Fig.~\ref{fig:errors} clearly illustrates that with the LHC
Run2 and Run3 statistics the effect of the anomalous $q\bar q$
background and that of an anomalous Higgs width cannot be disentangled
even for the dipole-like operator. \smallskip

In order to estimate the quantitative effect on the determination of the
Higgs width,  we perform a fit to the invariant mass event distribution
to extract the coefficient $X$ under the assumption that Nature
contains a SM Higgs but it also has anomalous $Z\bar{q}q$
couplings. So in each bin $j$, $N_{data}^j$ follows the predicted number of
events in Eq.~\eqref{eq:ndata}. We then fit that data to a model with
SM background and a Higgs signal with a width to be determined, {\em
  i.e.} we use $N^j_{model}(X)$ as Eq.~\eqref{eq:nx}. \smallskip

With these numbers of expected and data events we construct a
binned-likelihood $\chi^2$ for the four-lepton invariant mass
distribution using 100 GeV bins that start at 300 GeV:
\begin{equation}
  \chi^2(X) = 2\, \min_{\xi} \left\{\sum_{j=bins} \Big[
    (1+\xi)\, N^j_{model}(X)-N_{data}^j  + N_{data}^j\ln \frac{
      N_{data}^j}{(1+\xi)\, N^j_{model}(X)}\Big] +
    \frac{\xi^2}{\delta_\xi^2}\right\} \; .
\label{eq:chi2}
\end{equation}
In constructing Eq.~\eqref{eq:chi2} we have introduced the $\xi$ pull
to account for the possible effect of systematic uncertainties in the
overall normalization and to have a rough estimate of the impact of
these uncertainties. The LHC Run 2 luminosity uncertainty is of the
order of $2.5$\% and it is expected to be reduced to 1\% for the
HL-LHC~\cite{Dainese:2019rgk}. In addition, the LHC collaborations
estimate that the uncertainty due to the choice of the QCD scales
amounts to $\simeq 3-5$\%~\cite{Aaboud:2018puo,
  Sirunyan:2019twz}. \smallskip

It is important to stress that the analysis performed by the
experimental collaborations to extract the Higgs width make use of
much more sophisticated variables~\cite{ Aad:2015xua, Aaboud:2018puo,
  Khachatryan:2016ctc, Sirunyan:2019twz} which cannot be implemented
in our simplified simulation. Still, the most sensitive physics
variable to the Higgs width is the 4-lepton invariant mass dependence
of the different contributions. So, even if it is clear that our study
cannot reproduce the absolute precision in the width determination of
the experimental analysis, it should be good enough to quantify its
{\sl relative} change due to the presence of the anomalous $Z\bar{q}q$
couplings. \smallskip
  
\begin{figure}[h!]
\centering
 \includegraphics[width=0.75\textwidth]{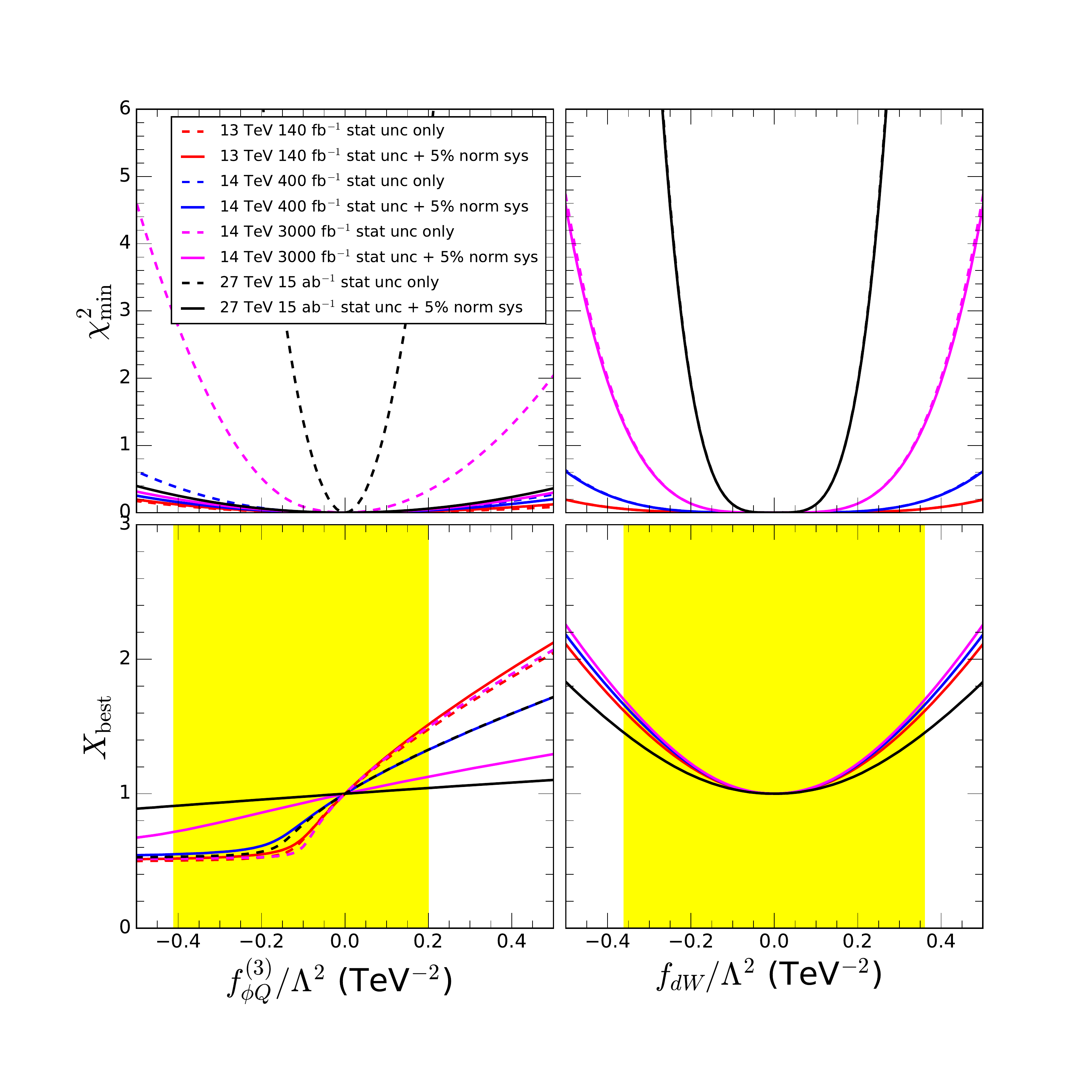}
 \caption{The upper (lower) panels display the minimum $\chi^2$ (best
   fit $X$) as a function of the Wilson coefficient
   $f^{(3)}_{\phi Q}/\Lambda^2$ (left) and
   $f^{(3)}_{dW}/\Lambda^2$(right).  The presently allowed regions for
   these couplings are indicated by the yellow bands. The different
   center-of-mass energies and integrated luminosities follow the
   color code indicated in the figure.  The dashed lines correspond to
   fits with no normalization uncertainty while the full lines
   correspond to the fit including a normalization uncertainty
   $\delta_\xi = 5$\%.}
  \label{fig:chix}
\end{figure}

The dependence of the quality of the fit with the value of the
$Z\bar{q}q$ anomalous coupling used in the generation of the data is
shown in Fig.~\ref{fig:chix}.  The left upper panel displays the
minimum $\chi^2$ as a function of the $f^{(3)}_{\phi Q}/\Lambda^2$ for
the different setups.  For each setup we show the results for two
analyses, one in which only statistical uncertainties are considered
(dashed lines) and another including the effect of an overall
normalization systematic (full lines).  As we can see, in the
unrealistic case of only having statistical errors, the quality of the
fit is not good for the HL-LHC and HE-LHC unless the anomalous
coupling is rather small. However, when the systematic normalization
error is included we find a satisfactory fit for all range of
anomalous couplings considered. \smallskip

In the left lower panel of Fig.~\ref{fig:chix} we plot the value of
the corresponding $X$ coefficient which leads to the best fit
($X_{best}$) for the analysis including the normalization systematic
uncertainty.  Obviously, as the data is assumed to contain always a SM
Higgs, the best fit for zero anomalous $Z\bar{q}q$ couplings is always
at $X_{best}=1$.  But, as seen in this figure, for the LHC Run2 (Run3)
$X_{best}$ deviates considerably from the SM value $X=1$ when the
anomalous $Z\bar{q}q$ contribution is included. In fact, this is the
case in a sizable fraction of the allowed range of this anomalous
coupling. This suggests that the limits on $\Gamma_H$ are affected in
the presence of the anomalous interactions while the global quality of
the fit is not.  On the contrary, as the integrated luminosity
increases for the HL-LHC and HE-LHC, $X_{best}$ remains around the SM
value for all presently allowed values of the anomalous coupling.
We notice in passing that the shift induced in $X_{best}$ from the SM
value of 1 is negative (positive) for negative (positive) value of
$f^{(3)}_{\phi Q}/\Lambda^2$. This is expected as the dominant
$1/\Lambda^2$ contribution is that induced on the $q\bar q$
background, that has the same sign as $f^{(3)}_{\phi Q}/\Lambda^2$. On
the other hand the contribution due to the gluon-gluon processes,
which has the opposite sign, is always subdominant; see dashed lines
in middle panels of Fig.~\ref{fig:m4l}. \smallskip

The corresponding results for the dipole operator coupling
$f^{(3)}_{dW}/\Lambda^2$ are shown in the right panels of
Fig.~\ref{fig:chix}. In this case we find that the inclusion of the
systematic normalization uncertainty has no effect in the quality of
the fit. This is so because the invariant mass dependence of the
anomalous $Z\bar{q}q$ events is sufficiently different from those due
to the Higgs exchange processes; see upper panels in
Fig.~\ref{fig:m4l}.  Consequently, the change induced in the $m_{4l}$
distribution of the number of events cannot be fitted with a modified
$X$ coefficient even if the prediction in all bins could be globally
shifted by an overall normalization factor.  The dependence of
$X_{best}$ upon $f^{(3)}_{dW}/\Lambda^2$ is shown in the right lower
panel of this figure from which we see, that for couplings close to
the limits of the presently allowed region, $X_{best}$ also deviates
noticeably from SM value of 1.  Notwithstanding, the effect for the
HL-LHC and HE-LHC is only important for values of these couplings for
which the fit is never good.  Finally, as expected, in this case the
shift in $X_{best}$ with respect to 1 is always positive irrespective
of the sign of the coupling because the anomalous contribution is
positive and quadratic in $f^{(3)}_{dW}/\Lambda^2$. \smallskip

\begin{figure}[h!]
\centering
 \includegraphics[width=0.9\textwidth]{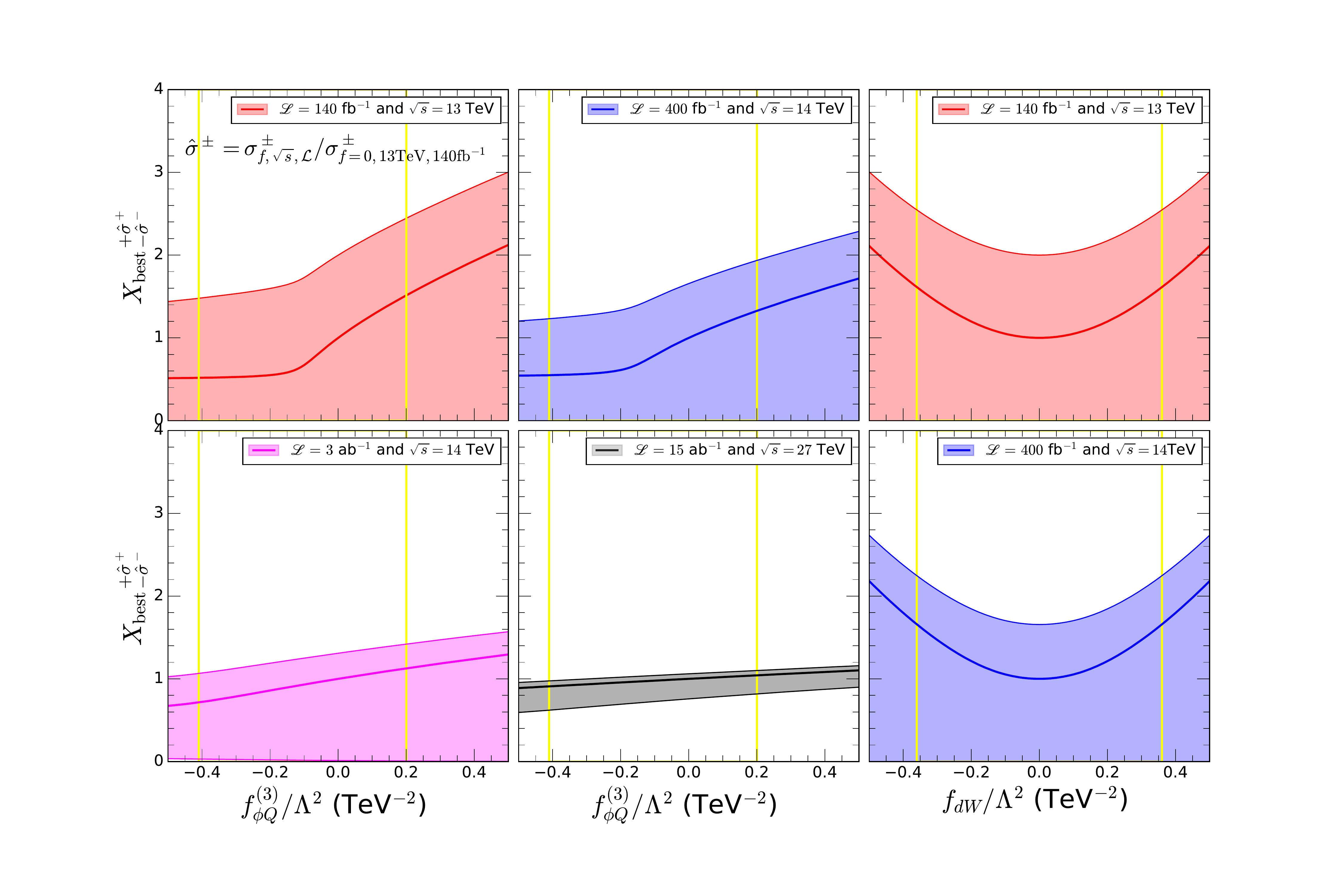}
 \caption{Dependence of the normalized 1$\sigma$ range of the
   coefficient $X$ with the Wilson coefficients
   $f^{(3)}_{\phi Q}/\Lambda^2$ (left/middle panels) and
   $f^{(3)}_{dW}/\Lambda^2$ (right panels); see text for details.  The
   presently allowed values for these couplings is indicated by the
   yellow vertical lines and the results are shown for different LHC
   setups as labeled in the figure.  In all cases the fit includes a
   normalization uncertainty $\delta_\xi = 5$\%.}
  \label{fig:1sig}
\end{figure}

Finally we study the impact of $Z$ anomalous coupling on
the apparent precision of the measurement of the Higgs width.  As
mentioned above with our simplified analysis we cannot reproduce the
precision attainable in dedicated experimental studies. But we can
still estimate how the presence of the anomalous $Z\bar{q}q$ coupling
affects that precision.  In the same fashion we can estimate how the
induced modification on the precision is expected to change in the
different setups. \smallskip

To this end we introduce {\sl normalized} $1\sigma$ upper and lower
errors as
\[
  \hat{\sigma}^\pm \equiv \frac{\sigma^\pm( f, \sqrt{s}, {\cal L})}
  {\sigma^\pm(0,  13 \hbox{ TeV}, 140 \hbox{ fb}^{-1}) } \;.
\]
Defined this way the extracted 68\% C.L. allowed range of the
coefficient $X$ for SM $Z\bar{q}q$ couplings in the Run 2 setup is
$X=1\pm 1$ by construction.  For the Run 2 setup, the variation of this
range with the anomalous coupling quantifies the relative change on
the determination of the central value and precision of Higgs width
due to the presence of the anomalous $Z\bar{q}q$ coupling. For other
setups it also accounts for the improvement in the determination of
$X$ due to the higher luminosity and center-of-mass energy. \smallskip

This is what we show in Fig.~\ref{fig:1sig} where we plot the
1$\hat\sigma$ allowed range around $X_{best}$ as a function of the
Wilson coefficients of the operators ${\cal O}_{\phi Q}^{(3)}$ and
${\cal O}_{dW}$ for the several setups for which we have found a good
fit in the presence of these interactions.  From this figure we see
that besides a shift in the best fit value, the presence of the
anomalous $Z\bar{q}q$ couplings results into a variation in the
attainable precision. \smallskip

For $f^{(3)}_{\phi Q}/\Lambda^2$ the presence of the anomalous
$Z\bar{q}q$ coupling leads to a noticeable increase (reduction) in the
uncertainty in the $X$ determination for positive (negative) values of
this Wilson coefficient. The effect is largest at the LHC Run 2 setup
(upper left panel).  At the LHC Run 3, see upper central panel, the
precision increases due to the larger statistics but, still, both the
best fit $X$ value and its error show a non-negligible dependence on
$f^{(3)}_{\phi Q}/\Lambda^2$.  However, for the HL-LHC and HE-LHC
setups (lower left and central panels), $\hat{\sigma}^\pm$ diminishes
considerably and it becomes rather independent of the $Z$ anomalous
coupling.  In other words, the foreseen accumulated luminosity for
these setups will allow for the measurement of the Higgs width which
will be robust under the possible presence of unaccounted anomalous
$Z\bar{q}q$ couplings. \smallskip

The right panels in Fig.~\ref{fig:1sig} show the corresponding results
for the dipole operator ${\cal O}_{dW}$ for the LHC Run 2 and Run 3
setups.  As seen in the figure the presence of a non-vanishing value of
this anomalous $Z\bar{q}q$ coupling always results into an apparent
less precise determination of $X$. \smallskip

\section{Conclusions}

In the Standard Model, the Higgs boson is a relatively narrow
resonance.  At the LHC, measuring its width by profiling the direct
production cross section is challenging. At present the most precise
available method to directly determine the Higgs boson width at LHC is
based on the study of the off-shell Higgs production in
$ p p \to Z Z$.  This analysis makes use of the fact that the Higgs
contribution, $g g \to H \to ZZ$, the continuum $g g \to Z Z$
generated via box diagrams and their interference depend on different
powers of the Higgs decay width; see Eq.~\eqref{eq:X}.  The
sensitivity of this process to $\Gamma_H$ is helped by the fact that,
in the Standard Model, the off-shell Higgs amplitude and its
interference with the continuum one give contributions of comparable
size but opposite sign, so they cancel up to an amount which depends
on the value of the width. For the same reason the study is sensitive
to the presence of new physics which modify the relative size of these
terms and as such it has been exploited in the literature. \smallskip

In this work we have focused on a different aspect of this process
which is associated to the presence of the irreducible background
originating from $q\bar q$ annihilation at tree level.  Being a tree
level process, this background is large and, therefore, any new
physics contribution to $Z\bar{q}q$ couplings, even if subdominant,
can be of comparable size to that due to corrections to gluon-gluon
Higgs terms from deviations of the Higgs width from the Standard Model
value. \smallskip

To quantify this possibility we have worked in the framework of
effective lagrangians parametrizing the effects of new physics on the
$Z$ couplings to quarks at low energies by the dimension-six effective
operators in Eqs.~\eqref{eq:hffop} and ~\eqref{eq:dipole}. In
Figs.~\ref{fig:m4l}, and~\ref{fig:errors} we explicitly show that,
despite the precise determination of the electroweak vector boson
couplings to fermions from electroweak precision data, the induced
anomalous fermionic couplings can give contributions to the $q\bar q$
background large enough to potentially affect the Higgs width
extraction from four-lepton events. \smallskip

In order to estimate the possible quantitative effect of these
anomalous background contributions, we have performed a statistical
analysis of the four-lepton invariant mass distribution.  Our
assumption is that the data contain a Standard Model Higgs plus a $Z$
boson with such anomalous couplings, but that the model to be fitted
has standard $Z\bar{q}q$ couplings and a Higgs with an unknown width
to be determined.  We have performed such analyses for the LHC Runs 2
and 3, as well as, for the high-luminosity LHC and high-energy
LHC. Our results are shown in Figs.~\ref{fig:chix} and
\ref{fig:1sig}. \smallskip

We conclude that the presence of anomalous $Z\bar{q}q$ contributions
to the background in the data which is not accounted for in the model
to be fitted: a) affects the apparent precision in the Higgs width
determination in the LHC Runs 2 and 3, b) induces a a shift in its
derived best fit value. Our results also show that the expected larger
integrated luminosities at the HL-LHC and HE-LHC runs should be enough
to mitigate these effects, making the determination of the Higgs width
in those setups robust under the presence of anomalous $Z$ couplings
within its present bounds, even if they are ignored when performing
the analysis. \smallskip

\section*{Acknowledgments}

We thank Nuno Rosa Agostinho for his contribution in the early stages
of this work.  O.J.P.E. is supported in part by Conselho Nacional de
Desenvolvimento Cent\'{\i}fico e Tecnol\'ogico (CNPq) and by
Funda\c{c}\~ao de Amparo \`a Pesquisa do Estado de S\~ao Paulo
(FAPESP) grant 2019/04837-9; E.S.A. thanks FAPESP for its support
(grant 2018/16921-1).
M.C.G-G is supported by  USA-NSF grant
PHY-1915093, by the spanish grants FPA2016-76005-C2-1-P, and by
AGAUR (Generalitat de Catalunya) grant 2017-SGR-929.

%
\bibliography{references}

\end{document}